\documentclass[12pt]{article}

\usepackage{geometry}

\usepackage{graphicx,amsfonts,bm,epsfig,color,latexsym}

\usepackage{amsmath}
\usepackage{amsfonts}
\usepackage{amssymb}
\usepackage{epstopdf}
\usepackage{subfig}
\usepackage{float}
\definecolor{background-color}{gray}{0.98}

\newcommand{\bea}{\begin{eqnarray}}
	\newcommand{\eea}{\end{eqnarray}}
\newcommand{\bes}{\begin{subequations}}
	\newcommand{\ees}{\end{subequations}}

\newcommand{\bd}{\begin{document}}
	\newcommand{\ed}{\end{document}}
\newcommand{\bc}{\begin{center}}
	\newcommand{\ec}{\end{center}}
\newcommand{\bfr}{\begin{flushright}}
	\newcommand{\efr}{\end{flushright}}
\newcommand{\lt}{\left}
\newcommand{\rt}{\right}
\newcommand{\vs}{\vspace}
\newcommand{\hs}{\hspace}
\newcommand{\beq}{\begin{equation}}
	\newcommand{\eeq}{\end{equation}}
\newcommand{\lb}{\linebreak}
\newcommand{\pb}{\pagebreak}
\newcommand{\mb}{\makebox}
\newcommand{\fb}{\framebox}
\newcommand{\mc}{\multicolumn}
\newcommand{\ben}{\begin{enumerate}}
	\newcommand{\een}{\end{enumerate}}
\newcommand{\bit}{\begin{itemize}}
	\newcommand{\eit}{\end{itemize}}
\newcommand{\oln}{\overline}
\newcommand{\un}{\underline}
\newcommand{\lefq}{\lefteqn}
\newcommand{\ba}{\begin{array}}
	\newcommand{\ea}{\end{array}}
\newcommand{\beqa}{\begin{eqnarray}}
	\newcommand{\eeqa}{\end{eqnarray}}
\newcommand{\beqas}{\begin{eqnarray*}}
	\newcommand{\eeqas}{\end{eqnarray*}}
\newcommand{\bfg}{\begin{figure}}
	\newcommand{\efg}{\end{figure}}
\newcommand{\bds}{\begin{displaymath}}
	\newcommand{\eds}{\end{displaymath}}
\newcommand{\btb}{\begin{tabbing}}
	\newcommand{\etb}{\end{tabbing}}
\newcommand{\para}{\parallel}
\newcommand{\pad}{\partial}
\newcommand{\nn}{\nonumber}
\newcommand{\la}{\leftarrow}
\newcommand{\ra}{\rightarrow}
\newcommand{\lgla}{\longleftarrow}
\newcommand{\lgra}{\longrightarrow}
\newcommand{\La}{\Leftarrow}\newcommand{\Ra}{\Rightarrow}
\newcommand{\Lra}{\Leftrightarrow}
\newcommand{\Lgla}{\Longleftarrow}
\newcommand{\Lgra}{\Longrightarrow}
\newcommand{\lan}{\langle}
\newcommand{\ran}{\rangle}
\renewcommand{\a}{\alpha}
\renewcommand{\b}{\beta}
\newcommand{\g}{\gamma}
\newcommand{\G}{\Gamma}
\renewcommand{\d}{\delta}
\newcommand{\eps}{\epsilon}
\newcommand{\Th}{\Theta}
\newcommand{\s}{\sigma}
\newcommand{\lam}{\lambda}
\newcommand{\D}{\Delta}
\newcommand{\ds}{\displaystyle}
\newcommand{\vare}{E}
\newcommand{\pr}{\prime}
\newcommand{\ro}{\rho}
\newcommand{\nab}{\nabla}
\newcommand{\m}{\mu}
\newcommand{\n}{\nu}
\newcommand{\Sg}{\Sigma}
\newcommand{\p}{\pi}
\newcommand{\R}{I\!\!R}
\newcommand{\om}{\omega}
\newcommand{\Om}{\Omega}
\newcommand{\ovra}{\overrightarrow}
\newcommand{\ze}{\zeta}
\newcommand{\vart}{\vartheta}
\newcommand{\tri}{\triangle}
\newcommand{\f}{\frac}
\newcommand{\iny}{\infty}
\newcommand{\pro}{\propto}
\renewcommand{\arraystretch}{1.25}


\title{Ro-vibrational energy and thermodynamic properties of molecules subjected to Deng-Fan potential through an improved approximation}

\author{Debraj Nath \thanks{Department of Mathematics, Vivekananda College, Thakurpukur, Kolkata-700063, India. Email: debrajn@gmail.com}, Amlan K.~Roy\thanks{Department of Chemical Sciences, Indian Institute of Science Education and Research (IISER) Kolkata, Nadia, Mohanpur-741246, WB, India. Email: akroy@iiserkol.ac.in, akroy6k@gmail.com.}
\thanks{AKR dedicates this article to the loving memory of his kind-hearted teacher, late Prof.~Hrishikesh Pradhan, P.~K.~College, Contai. Through his 
relentless, selfless service, he has influenced and inspired many chemists, over the years.}}

\begin{document}
		
		\maketitle

\begin{abstract}
Accurate solution of the Schr\"odinger equation with Deng-Fan potential is presented by means of Nikiforov-Uvarov method. A modified 
Pekeris-type approximation is proposed for the centrifugal term, from a linear combination of the $r \to 0$ and $r \to r_e$ limits. It can 
potentially offer a series of approximations (depending on an 
adjustable parameter $\lambda$). The existing approximations in the literature can then be recovered in certain special cases. Its efficiency 
and feasibility is demonstrated by a critical comparison of eigenvalues produced at various $\lambda$'s for four molecules, \emph{viz.}, 
H$_2$, LiH, HCl and CO. Analytical expressions are 
derived for energies, eigenfunctions and the thermodynamic properties such as vibrational mean free energy, vibrational free energy, 
vibrational entropy and vibrational specific heat. The effect of quantum correction on partition function and thermodynamic properties 
is discussed by including the correction up to 10th-order, for H$_2$ and LiH. The effect of $\lambda$ parameter on these properties is 
also studied.

\end{abstract}

\clearpage

\makeatletter
\renewcommand\@biblabel[1]{#1.}
\makeatother

\bibliographystyle{apsrev}

\renewcommand{\baselinestretch}{1.5}
\normalsize

\clearpage
\section{Introduction}
A correct and proper representation of potential energy as a function of internuclear distance is an essential and indispensable tool for accurate understanding of structure and dynamics of a molecular system. An ideal potential energy function should show proper limiting behavior, i.e., $V(0)= \infty$ and approach a constant at infinite distance. Analytical eigenvalues and eigenfunctions are desirable, as they facilitate accurate estimation of transition frequencies and matrix elements. The oldest and arguably the most celebrated empirical Morse potential \cite{morse29} was suggested almost ninety years ago. This widely used simple analytical functional has found valuable applications \cite{rong2003} in several branches in physics and chemistry including ro-vibrational states of diatomic molecules, adsorption of atoms/molecules on solid surface, deformation of cubic metals etc. The three-parameter exponential function is represented by, 
\beq
U(r)=D_e \left[ 1-\exp^{-\alpha (r-r_e)} \right]^2, 
\eeq 
where $D_e$ denotes dissociation energy, $r_e$ corresponds to equilibrium internuclear distance and the Morse parameter $\alpha$ is defined in terms of force constant $k_e$ as $\alpha=\sqrt{k_e/2D_e}$. Despite its remarkable success, the potential has a finite value, as internuclear distance approaches zero, and does not lead to the expected infinity nature. Thus when employed for the vibrational levels of diatomic molecule, the Morse oscillator shows considerable deviation from experimentally observed values. There has been many attempts to address these issues leading to a variety of proposals in the literature over the years. Some of the most prominent and notable ones are listed as follows: Kratzer \cite{kratzer20,bayrak2007,ikhdair2009}, Rosen-Morse \cite{rosen32}, Manning-Rosen \cite{manning33,chen2009,nasser2013, roy2014a}, P\"oschl-Teller \cite{poschl33,dong2002,aktas2004}, Linnett \cite{linnett40}, Hulth\'en \cite{hulthen42,roy2005,jia2009,gu2010}, Lippincott \cite{lippincott53}, shifted/Deng-Fan (DF) \cite{deng57,rong2003,oyewumi2013,roy2014b,oluwadare2018,vogt2018,boukabcha2018}, Tietz-Hua \cite{tietz63,hua90,hamzavi2012,roy2014}, Schl\"oberg \cite{schloberg86}, Zavitsas \cite{zavitsas91}, deformed Rosen-Morse \cite{egrifes99, sun2013,jia2013}, Woods-Saxon \cite{fakhri2004,guo2005,ikhdair2010}, pseudoharmonic \cite{ikhdair2006,oyewumi2008}, Hajigeorgiou \cite{haji2010}. 

The current communication deals with the DF potential for diatomic molecules put forth about 70 years ago. This three-parameter function given in the following expression
\begin{equation} \label{eq:2}
	V(r)=D_e \left( 1- \frac{b} {e^{a r}-1} \right)^2, \ \ \ \ b=e^{a r_e} -1, \ \ \ \  r \in (0,\infty), 
\end{equation}
has aroused considerable interest in last two decades. Here the three positive parameters $D_e, r_e, a$ represent dissociation energy, equilibrium internuclear distance and radius of the potential well respectively. As internuclear distance approaches the limiting values zero and infinity, it offers a qualitatively correct asymptotic nature. This is also often referred as Generalized Morse potential due to its qualitative resemblance to Morse potential \cite{mesa98,rong2003}. This is also connected to another important diatomic potential due to Manning-Rosen \cite{manning33}. 

Like the case of many other practical and realistic potentials, the relevant {\bf Schr\"odinger equation with centrifugal term} can not be solved analytically for eigenvalues and eigenfunctions. This has lead to the development of a number of {\bf theoretical methodologies of approximation to the centrifugal term} in both non-relativistic and relativistic case. Some of the prominent ones may be mentioned. The authors in \cite{mesa98} have discussed the exact solvability problem via an $SO(2,2)$ symmetry algebra. The eigenvalues and eigenfunctions for $\ell=0$ states were studied by an algebraic method \cite{codriansky99}. Approximate analytical solutions for rotating DF potential for arbitrary $n, \ell$ quantum numbers were given in terms of generalized hyper-geometric functions $_2F_1(a,b;c;z)$ \cite{dong2008}. An improved approximation scheme for the centrifugal term, along with a super-symmetric shape invariance approach  was employed in \cite{zhang2011}; analytical solution for Dirac equation has been presented with latter approach \cite{zhang2009}. Solutions of Klein-Gordon equation with spinless particle for $\ell \neq 0$ states were considered in \cite{dong2011}. A Pekeris approximation for centrifugal term and a Nikiforov-Uvarov (NU) method was employed for Dirac, Klein-Gordon and Schr\"odinger equation \cite{oluwadare2012,hamzavi2013}. Some other methods include a numerical generalized pseudospectral (GPS) scheme \cite{roy2014b}, exact quantization rule \cite{falaye2015,oluwadare2018}, a Feynman path integral formalism for D-dimensional problem \cite{boukabcha2018}, etc. A comparative analysis for the performance of Morse, Manning-Rosen, Schl\"oberg and DF potential in the context of diatomic molecules has been offered in \cite{wang2012}.

Our objective in this communication is two-fold. At first, we present accurate eigenvalues and eigenfunctions of the DF potential by means of a simple novel approximation for the centrifugal term, within the framework of NU method. This is accomplished by taking a combination of approximations at $r \to 0$ and $r \to r_e$ limits. Its performance is critically analyzed by presenting the eigenvalues at various values of tunable parameter. These are compared with existing approximations available in the literature. The calculated ro-vibrational eigenvalues are offered for two representative diatomic molecules (H$_2$, LiH). Both circular ($s$) as well as non-zero-$\ell$ states are obtained accurately. Then in the second stage, analytical expressions for vibrational partition function and related thermodynamic properties such as internal energy, free energy, entropy, specific heat capacity are derived. They are examined in terms of (i) temperature and (ii) the impact of quantum correction on these. 
The article is organized as follows. Section~2.1 provides the relevant details pertaining to the new approximation. The necessary expressions for eigenvalues and eigenfunctions within the NU method are detailed in Sec.~2.2. The associated energy spectrum is analyzed in Sec.~2.3. The expressions for thermodynamical quantities are given and discussed in Sec.~3 for a representative set of four diatomic molecules (H$_2$, LiH, HCl and NO). Additionally the effect of quantum correction on the thermal quantities are discussed as well. Finally a few comments are made in Sec.~4.  

\section{The methodology} 
\subsection{Analytical solutions of DF potential with centrifugal term within a new approximation}
\begin{figure}
	\includegraphics[width=16cm,height=6cm]{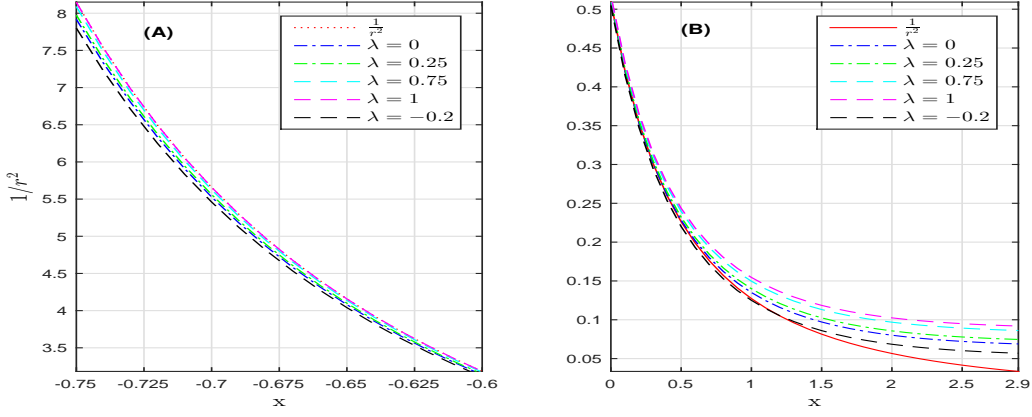}
	\caption{\label{Fig.1approximation} Plot of the centrifugal term $\f{1}{r^2}$ and the three approximations given in Eqs.~(\ref{series.x}), (\ref{greene.app}) and (\ref{comvex}) for H$_2$ molecule. The two situations $\lam = 0,1$ correspond to Eqs.~(\ref{series.x}) and (\ref{greene.app}) respectively. Panels (A), (B) refer to $(-)$ ve and $(+)$ ve regions of $x$.} 
\end{figure}

The time-independent non-relativistic Schr\"odinger equation for a diatomic molecule in presence of DF potential in Eq.~(\ref{eq:2}), can be 
written as, 
\beq
-\f{\hbar^2}{2\mu}\nabla^2\psi ({\bf r}) +V({\bf r})\psi ({\bf r}) =E\psi ({\bf r}) , 
\eeq 
where
\beq
\nabla^2=\f{1}{r^2}\f{\partial}{\partial r}\left(r^2\f{\partial}{\partial r}\right)+\f{1}{r^2\sin\theta}\f{\partial}{\partial 
	\theta}\left(\sin\theta\f{\partial}{\partial \theta}\right)+\f{1}{r^2\sin^2\theta}\f{\partial^2}{\partial \phi^2},
\eeq
with $\mu$ and $\hbar$ denoting the reduced mass 
and Planck constant respectively. 
Let us define, 
\beq
\psi({\bf r})=\f{1}{\sqrt{2\pi}}\f{R(r)}{r}Y_l^m(\theta)\,e^{im_l\phi},
\eeq 
where $R(r)$ and $Y_l^{m_l}$ signify the usual radial and angular functions, while $l,m_l$ represent the angular quantum numbers. Then one obtains 
the following radial Schr\"odinger equation, 
\beq\label{eq:6}
\f{d^2R (r) }{dr^2}+\left[\f{2\mu}{\hbar^2}E-\f{2\mu }{\hbar^2}V(r)-\f{l(l+1)}{r^2}\right]R (r) =0,
\eeq 
where $l(l+1)$ is the separation constant. Now one can expand (Pekeris-type approximation) $\f{1}{r^2}$ as  below, 
\beq\label{series.x}
\ba{ll}
\f{1}{r^2}&=\f{f(x)}{r_e^2}=\f{1}{r_e^2}\ds\sum\limits_{n=0}^{\infty}\f{x^n}{n!}\left[\f{d^nf(x)}{dx^n}\right]_{x=0}\approx 
\f{1}{r_e^2}\ds\sum\limits_{n=0}^{2}\f{x^n}{n!}\left[\f{d^nf(x)}{dx^n}\right]_{x=0}+O(x^3),
\ea
\eeq 
where  
\beq
f(x)=c_0+\f{c_1\,s}{1-s}+\f{c_2\,s^2}{\left(1-s\right)^2},
\eeq
and
\beq\label{transformation}
s=e^{-\a r}, r=r_e(1+x), 
\eeq 
{\bf where $x$ is a dimensionless variable.}
The coefficients $c_0,c_1$ and $c_2$ are dimensionless parameters. In order to find these coefficients for non-zero $l$, one may express 
$\f{1}{r^2}$ as \cite{pekeris34,hua90,badawi72}, 
\beq\label{order.x3}
\ba{ll}
\f{1}{r^2}&=\f{1}{r_e^2}\ds\sum\limits_{n=0}^{\infty}(-1)^n(n+1)x^n,\\
&\approx\f{1}{r_e^2}\left(1-2x+3x^2+O(x^3)\right).
\ea 
\eeq  
Comparing the coefficients of Eq.~(\ref{series.x}) and Eq.~(\ref{order.x3}), \cite{badawi72,mustafa2015,boukabcha2018}, one obtains, 
\beq
\ba{ll}
c_0&=\f{1}{u^2}\left[3-3u+u^2+\left(2u-6\right)s_e+\left(u+3\right)s_e^2\right],\\
c_1&=\f{2}{s_eu^2}\left(1-s_e\right)^2\left((3+u)s_e+2u-3\right), \\
c_2&=-\f{1}{s_e^2u^2}\left(1-s_e\right)^3\left((3+u)s_e+u-3\right),\\
s_e&=e^{-u},~u=\a r_e.
\ea
\eeq 
Let us now consider another approximation (Greene-type) \cite{greene76,oluwadare2018,oyewumi2013a,oyewumi2013,hamzavi2013} of the form, 
\beq
\ba{ll}\label{greene.app}
\f{1}{r^2}&\approx\a^2\left(\f{1}{12}+\f{s}{1-s}+\f{s^2}{(1-s)^2}\right),\\
\ea 
\eeq 
{\bf If we express $\a^2\left(\f{1}{12}+\f{s}{1-s}+\f{s^2}{(1-s)^2}\right)$ in the explicit form of $r$, then we obtain $\a^2\left(\f{1}{12}+\f{s}{1-s}+\f{s^2}{(1-s)^2}\right)\approx\f{1}{r^2}+\a^2\left(\f{(\a r)^2}{240}-\f{(\a r)^4}{6048}+\f{(\a r)^6}{172800}+O((\a r)^8)\right)$. Hence it} is better than Eq.~(\ref{series.x}), when $\a r\ll 1$. In the limiting case, 
\beq
\ba{ll}
\f{1}{r^2}&=\lim\limits_{\a\rightarrow 0}\a^2\left(\f{1}{12}+\f{s}{1-s}+\f{s^2}{(1-s)^2}\right).
\ea
\eeq 
Therefore, for a fixed $\a$, Eq.~(\ref{greene.app}) is a good approximation near $x=-1$ i.e., around $r=0$ and Eq.~(\ref{series.x}) 
is good near $x=0$ i.e., in region close to $r=r_e$. This prompts us to consider a linear combination of above two approximations, as in the 
following, 
\beq\label{comvex}
\f{1}{r^2}\approx\a^2\left(d_0(\lam)+\f{d_1(\lambda)\,s}{1-s}+\f{d_2(\lambda)\,s^2}{(1-s)^2}\right),
\eeq 
where
\beq
d_0(\lam)=\f{\lam}{12}+\f{(1-\lam)c_0}{u^2},~d_1(\lam)=\lam+\f{(1-\lam)c_1}{u^2},~d_2(\lam)=\lam+\f{(1-\lam)c_2}{u^2}.
\eeq 
From the above equation, it is observed that, $\lam=0$ implies Eq.~(\ref{series.x}) and $\lam=1$ signifies Eq.~(\ref{greene.app}). Therefore, one can 
use Eq.~(\ref{comvex}) as a more general form of approximation for the centrifugal term $\f{1}{r^2}$. As $r$ increases, we can consider Eq.~(\ref{comvex}) 
for negative value of $\lam$. In Fig.~\ref{Fig.1approximation}, the centrifugal term $\f{1}{r^2}$ is plotted along with three approximations 
defined in Eqs.~(\ref{series.x}), (\ref{greene.app}) and (\ref{comvex}), for H$_2$. The panels (A), (B) refer to the corresponding functions 
in $(-)$ ve and $(+)$ ve domains of $x$. {\bf For $0\le \lambda \le 1$, the approximation of Eq.~(\ref{comvex}) is a convex combination of 
Eqs.~(\ref{series.x}) and (\ref{greene.app}), and the corresponding centrifugal term is a good approximation for $0\le r\le r_e$. For negative values of $\lambda$, Eq.~(\ref{comvex}) is a good approximation to the centrifugal term for $r>r_e$, which is clear from 
Fig.~\ref{Fig.1approximation}(B).} It is noticed that the approximation defined in 
Eq.~(\ref{comvex}) is the best amongst them. This is generally found to be true for other molecules as well.

Using the transformation in Eq.~(\ref{transformation}) and approximation of Eq.~(\ref{comvex}) in Eq.~(\ref{eq:6}) we obtain, 
\beq
\ba{l}\label{Eq.R.NU}
\f{d^2R}{ds^2}+\f{1-s}{s(1-s)}\f{dR}{ds}+\f{-As^2+Bs-C}{s^2(1-s)^2}R=0,
\ea 
\eeq 
where
\beq\label{g0g1g2g}
\ba{ll}
A &=\bar{\varepsilon}^2+2b\b^2+b^2\b^2+\g(d_2-d_1),\\
B &=2\bar{\varepsilon}^2+2b\b^2-\g d_1,\\
C &=\bar{\varepsilon}^2,\\
\bar{\varepsilon}^2&=\b^2+\g d_0-\varepsilon, \\
\b^2&=\f{2\mu D_e}{\a^2\hbar^2},\\
\varepsilon&=\f{2\mu E}{\a^2\hbar^2},\\
\g&=l(l+1).
\ea
\eeq 
Therefore, it reduces to an energy-dependent problem of a second-order differential equation. In the next subsection, we provide the solution of 
Eq.~(\ref{Eq.R.NU}) by making use of NU method.

\subsection{Nikiforov-Uvarov Method}\label{NUmethod}
One can compare Eq.~(\ref{Eq.R.NU}) with the following Eq. \cite{nikiforov88,berkdemir2005,tezcan2009,oyewumi2013,udoh2019}, 
\beq\label{Eq.NU}
\f{d^2R}{ds^2}+\f{\widetilde{\tau}(s)}{\sigma(s)}\f{dR}{ds}+\f{\widetilde{\sigma}(s)}{\sigma^2(s)}R=0,
\eeq
where
\beq
\ba{ll}
\widetilde{\tau}(s)&=1-s,~\sigma(s)=s(1-s),~
\widetilde{\sigma}(s)
=-A\,s^2+B\,s-C.
\ea
\eeq
Let us consider the solution of Eq.~(\ref{Eq.NU}) in following form, 
\beq
R(s)=\phi(s)y(s).
\eeq
Then one may obtain \cite{nikiforov88}, 
\beq\label{Eq.y}
\sigma(s)y^{''}(s)+\tau(s)y'(s)+\nu\, y(s)=0,
\eeq
and
\beq
\ba{l}
\phi(s)=\ds e^{\int\f{\pi(s)}{\sigma(s)}ds},
\ea
\eeq 
where
\beq
\ba{ll}
\pi(s)&=\f{\sigma'(s)-\widetilde{\tau}(s)}{2}\pm\sqrt{\left(\f{\sigma'(s)-\widetilde{\tau}(s)}{2}\right)^2-\widetilde{\sigma}(s)+k\,\sigma(s)},\\
\tau(s)&=\widetilde{\tau}(s)+2\pi(s),\\
\nu&=k+\pi'(s).
\ea 
\eeq
Here $\nu$ and $k$ are real constants independent of $s$. Since $\pi(s)$ is a polynomial in $s$, $\left(\f{\sigma'(s)-\widetilde{\tau}(s)}{2}\right)^2-\widetilde{\sigma}(s)+k\,\sigma(s)$ is a square of a polynomial in $s$, and the solutions of Eq.~(\ref{Eq.y}) are given by,
\beq
y_n(s)=\f{1}{\rho(s)}\f{d^n}{ds^n}\left[\sigma^n(s)\rho(s)\right], 
\eeq
where 
\beq
\ba{l}
\rho(s)=\ds \left[\sigma(s)\right]^{-1}\,e^{\int\f{\tau(s)}{\sigma(s)}ds}.
\ea 
\eeq  
Therefore, the corresponding eigenvalues are obtained from the relation, 
\beq\label{eigen.nu}
\nu_n=-n\,\tau'(s)-\f{n(n-1)}{2}\sigma^{''}(s)=-n\,\tau'(s)+n(n-1),~n=0,1,2,\dots. 
\eeq 
According to NU method, we obtain a pair of possible values of $k$, which are given by, 
\beq
\ba{ll}\label{kpitaunu}
k_{\pm}
&=-\g d_1+2b\b^2\pm\bar{\varepsilon}\sqrt{4b^2\b^2+4\g d_2+1}.
\ea
\eeq
Now using Eq.~(\ref{kpitaunu}), we obtain a set of possible values of $\pi(s)$ as, 
\beq
\ba{ll}\label{pi.tau.nu}
\pi(s)
&=-\f{s}{2}\pm\left\{\ba{lll}\left(\bar{\varepsilon}-\f{1}{2}\sqrt{4b^2\b^2+4\g d_2+1}\right)s+\sqrt{C},&k=k_+,& k_+-B>0\\\left(\bar{\varepsilon}+\f{1}{2}\sqrt{4b^2\b^2+4\g d_2+1}\right)s+\sqrt{C},&k=k_-,& k_--B>0\\
\left(\bar{\varepsilon}-\f{1}{2}\sqrt{4b^2\b^2+4\g d_2+1}\right)s-\sqrt{C},&k=k_+,& k_+-B<0\\\left(\bar{\varepsilon}+\f{1}{2}\sqrt{4b^2\b^2+4\g d_2+1}\right)s-\sqrt{C},&k=k_-,& k_--B<0\ea \right\}.
\ea
\eeq
Since $\tau'(s)<0$, in this work, we have considered $k=k_-$, where $k_--B<0$ and select the following $\pi(s)$, 
\beq
\ba{ll}
\pi(s)
&=-\f{s}{2}-\left(\bar{\varepsilon}+\f{1}{2}\sqrt{4b^2\b^2+4\g d_2+1}\right)s+\sqrt{C}.
\ea
\eeq  
In that case, one gets,  
\beq
\ba{ll}\label{rho.phi}
\rho(s)&=s^{2\bar{\varepsilon}}(1-s)^{2L-1},\\
\phi(s)&=s^{\bar{\varepsilon}}(1-s)^L,
\ea 
\eeq
and
\beq
\ba{ll}
y_n&=s^{-2\bar{\varepsilon}}(1-s)^{-2L+1}\f{d^n}{dx^n}\left[s^{n+2\bar{\varepsilon}}(1-s)^{n+2L-1}\right]=n!\,P_n^{(2\bar{\varepsilon},2L-1)}(1-2s),\\
&=\f{\G\left(n+2\bar{\varepsilon}+1\right)}{\G\left(2\bar{\varepsilon}+1\right)}{}_2F_1\left(-n,n+2\bar{\varepsilon}+2L;2\bar{\varepsilon}+1;s\right),
\ea
\eeq 
where
\beq 
L=\f{1}{2}+\sqrt{\f{1}{4}+b^2\b^2+\g d_2}.
\eeq
The ro-vibrational energies $E_{nl}$ of the DF potential can then be derived from the relation, 
\beq\label{deri.tau}
k+(2n+1)\pi'(s)=n^2.
\eeq
Finally, the ro-vibrational energy spectrum can be secured from, 
\beq\label{Energy.En.mp}
\ba{ll}
E_{nl}
&=d_4-\f{\a^2\hbar^2}{2\mu}\left(\f{d_3^2}{(n+L)^2}+\f{(n+L)^2}{4}\right),
\ea 
\eeq
where
\beq
\ba{ll}
d_3&=\f{1}{2}(2+b)b\b^2+\f{1}{2}\g (d_2-d_1),\\
d_4&=D_e+\f{l(l+1)\hbar^2\a^2}{2\mu}d_0+\f{\a^2\hbar^2 d_3}{2\mu}.
\ea
\eeq  
Therefore, the radial solution can be expressed as, 
\beq
R_n(r)=N_{nl}\,s^{\bar{\varepsilon}}(1-s)^{L}\,{}_2F_1\left(-n,n+2\bar{\varepsilon}+2L;2\bar{\varepsilon}+1;s\right),
\eeq 
where 
\beq
N_{nl}=\left[\f{2\bar{\varepsilon}\a(n+\bar{\varepsilon}+L)\G(n+2\bar{\varepsilon}+1)\G(n+2\bar{\varepsilon}+2L)}{n!(n+\bar{\varepsilon})\G(n+2L)\left[\G(2\bar{\varepsilon}+1)\right]^2}\right]^{\f{1}{2}},
\eeq 
is the normalization constant, derived from, 
\beq
\ds\int|\psi_{n,l,m_l}({\bf r})|^2\,d{\bf r}=1,~ d{\bf r}=r^2\sin\theta\,dr\,d\theta\,d\phi. 
\eeq 
Finally, the explicit form of eigenfunctions of DF potential can be written as, 
\beq
\psi_{n,l,m_l}({\bf r})=\f{N_{nl}}{\sqrt{2\pi}}\f{R_n(r)}{r}Y_{l}^{m_l}(\theta)\,e^{im_l\phi},
\eeq
where
\beq\label{Ylm}
Y_l^{m_l}(\theta)=\left[\f{(2l+1)(l-|m_l|)!}{2(l+|m_l|)!}\right]^{\f{1}{2}}\,P_{l}^{|m_l|}(\cos\theta)\,e^{im_l\phi}.
\eeq 

\subsection{Energy spectrum}
To illustrate the effectiveness of our approximation, selected energies of DF potential are tabulated for two diatomic molecules, \emph{viz.}, 
H$_2$ (homo-nuclear) and LiH (hetero-nuclear). In each case, 10 states are considered to cover a broad range. The parameters used
are as follows: Bohr radius = 0.52917721092\AA, Hartree energy = 27.21138505 eV, electron rest mass = 5.48577990946$\times 10^{-4}$ amu. 
Spectroscopic parameters for H$_2$ are: $D_e= 4.7446$ eV, $r_e=0.7416$\AA, $\alpha=1.9426$\AA$^{-1}$, $\mu=0.50391$ amu, while for LiH, 
these are: $D_e= 2.515283695$ eV, $r_e=1.5956$\AA, $\alpha=1.1280$\AA$^{-1}$, $\mu=0.8801221$ amu. In columns 3-7 of Table~1, our estimated energies
from Eq.~(34) are presented for five $\lambda$, namely $-$0.25, 0, 0.75, 1 and 1.25 respectively. A total of 10 states are given for each molecule
for arbitrary $n,l$ quantum numbers. It may be noted that, the energies were considered by 
means of super-symmetric quantum mechanics \cite{mustafa2015} and path integral approach \cite{boukabcha2018}; these correspond to the case of
$\lambda=0$. On the other hand, the energies of NU method \cite{hamzavi2013,oyewumi2013}, asymptotic iteration \cite{oyewumi2013a} and Hellmann-Feynman 
theorem \cite{oluwadare2018} represent the $\lambda=1$ case. For the so-called circular or $l=0$ states, all the columns produce identical 
energies, which is expected from Eq.~(\ref{comvex}). But for $l \neq 0$ states, the energies in five columns differ from each other. It is noticed that
the energies in 6th column (relating to $\lambda=1$) completely reproduce those reported in \cite{oyewumi2013}. 

\begin{table}               
	\caption{\label{Table1.Energy} Selected ro-vibrational energies (in eV) of H$_2$ and LiH, represented by DF potential. 
		These correspond to five representative $\lambda$'s, indicated in parentheses. For more details, see text.}
		\begin{tabular}{lr|lllll}\hline
			$n$	& $l$ & $E_{nl}(-0.25)$	&$E_{nl}(0)$& $E_{nl}(0.75)$& $E_{nl}(1)$\footnotemark[1]	&	 $E_{nl}(1.25)$	\\   \hline
			& & \multicolumn{4}{c}{ H$_2$} &   \\  \hline
			
			0&    0&     0.349980221&      0.349980221&     0.349980221&     0.349980221&     0.349980221\\
			1    &0  &   0.996777053 &     0.996777053  &   0.996777053    & 0.996777053 &    0.996777053\\
			1    &1  &   1.009941717   &   1.010018022 &    1.010246935 &    1.010323238  &   1.010399541\\
			2    &0  &   1.580248366   &   1.580248366&     1.580248366 &    1.580248366    & 1.580248366\\
			2    &1  &   1.592275484  &    1.592360547&     1.592615732  &   1.592700793    &1.592785853\\
			2    &2  &   1.616260599&      1.616516418&     1.617283848&     1.617539648 &    1.617795443\\
			5    &0   &  2.986148433 &     2.986148433 &    2.986148433 &    2.986148433 &    2.986148433\\
			5    &5    & 3.118869192 &     3.120534519 &    3.12552957 &    3.127194276  &   3.128858827\\
			5    &10 &    3.452477485 &     3.458819802    & 3.477833081  &   3.484166293     &3.490497236\\
			5    &15 &    3.935487614    &  3.950184048  &   3.994199652 &    4.008847068  &   4.023482314\\\hline
			& & \multicolumn{4}{c}{ LiH} &\\ \hline
			0 &   0    & 0.103334650  &    0.103334650   &  0.103334650    & 0.103334650   &  0.103334650\\
			1   & 0&     0.302005955&      0.302005955&     0.302005955&     0.302005955&     0.302005955\\
			1 &   1 &    0.303739903 &     0.303759653  &   0.303818903 &    0.303838653&     0.303858402\\
			2 &   0  &   0.490685861  &    0.490685861  &   0.490685861  &   0.490685861 &    0.490685861\\
			2  &  1&     0.492346421 &     0.4923672887 &    0.492429891 &    0.492450759 &    0.492471626\\
			2  &  2 &    0.495665829 &     0.495728464 &    0.495916365 &    0.495978997&     0.496041629\\
			5    &0&     0.999006401 &     0.999006401 &    0.999006401 &    0.999006401&     0.999006401\\
			5 &   5  &   1.020697858   &   1.021060955  &   1.022150163   &  1.022513206  &   1.022876235\\
			5 &   10   &  1.077958316   &   1.079301598 &    1.083330322  &   1.084672857  &   1.086015205\\
			5  &  15   &  1.169212076  &    1.1721857 &    1.181101026 &    1.184070957  &   1.187039967	\\\hline
		\end{tabular} 
		\begin{tabbing}
			\footnotemark[1]{These energies coincide with those from \cite{oyewumi2013}.}  
		\end{tabbing} 
\end{table}

These observations of Table~1 are graphically shown in Fig.~2, where the ro-vibrational energies of H$_2$ molecule are depicted for three
representative states. The left and right panels give $D_e-E_{nl}$ and $E_{nl}$ respectively; the state labels are provided in parentheses. Shifted 
energies are plotted as well, to 
ease the facilitation of comparison with select literature values. Five values of $\lambda$ ranging from $-0.5$ to 1.5 including 0, are considered. 
In each case, a line is generated. In the right panel, only reference energies of \cite{oyewumi2013} are available, which are marked 
in blue color. For $\lambda=1$, they
fall on the straight line. In left panels, reference energies are available in scattered manner. But in any case, the available ones 
related to $\lambda=1$ directly fall on the line as expected. The GPS energies of \cite{roy2014b} remain slightly left off $\lambda=0$, in 
$l \neq0$ states in panels (B) and (F). However, this can happen as it is a purely numerical approach. A similar plot is also found for 
LiH molecule in Fig.~3. In this case, the reference results of \cite{oluwadare2018} falls slightly right off $\lambda=1$. This is presumably 
because of the differences in implementation of their method from ours; we use a NU scheme while they employed a Hellmann-Feynman approach. 
Note that in panel (C), the path-integral \cite{boukabcha2018} and GPS \cite{roy2014b} energies overlap. The latter results, like H$_2$, hover in the 
neighborhood of $\lambda=0$. For a few other molecules, quite similar observations are made; hence they are omitted here.  

\begin{figure}
	\includegraphics[width=18cm,height=16cm]{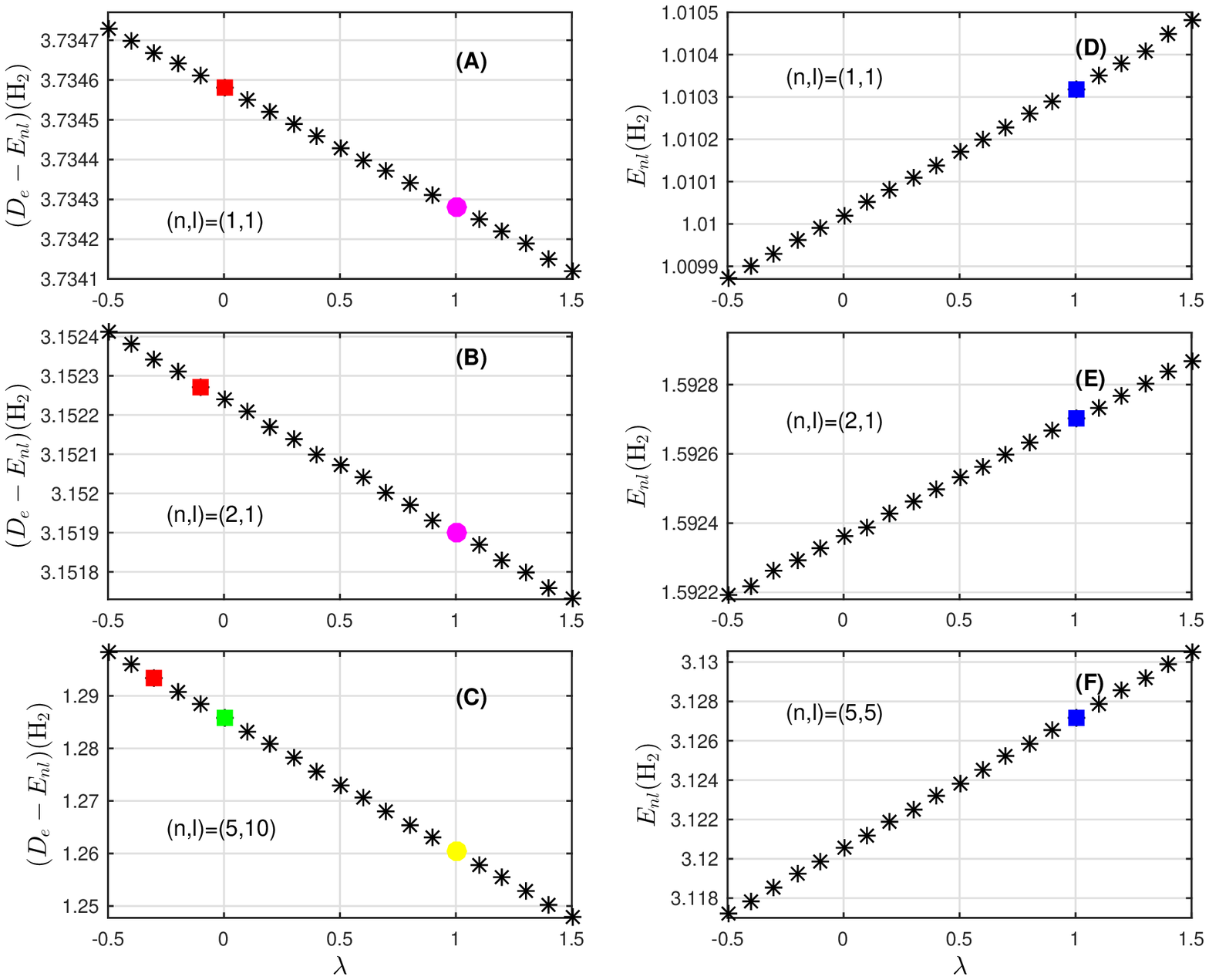}
	\caption{\label{Fig.2energy} Plot of $(D_e-E_{nl})$ (left) and $E_{nl}$ (right) of H$_2$ molecule in eV, against $\lambda$. The state indices are given inside the panels. The colored points show literature energies from various references as: red \cite{roy2014b}, magenta \cite{oluwadare2018}, green \cite{boukabcha2018}, yellow \cite{oyewumi2013a} and blue \cite{oyewumi2013}.}
\end{figure}

\begin{figure}
	\centering
	\includegraphics[width=18cm,height=16cm]{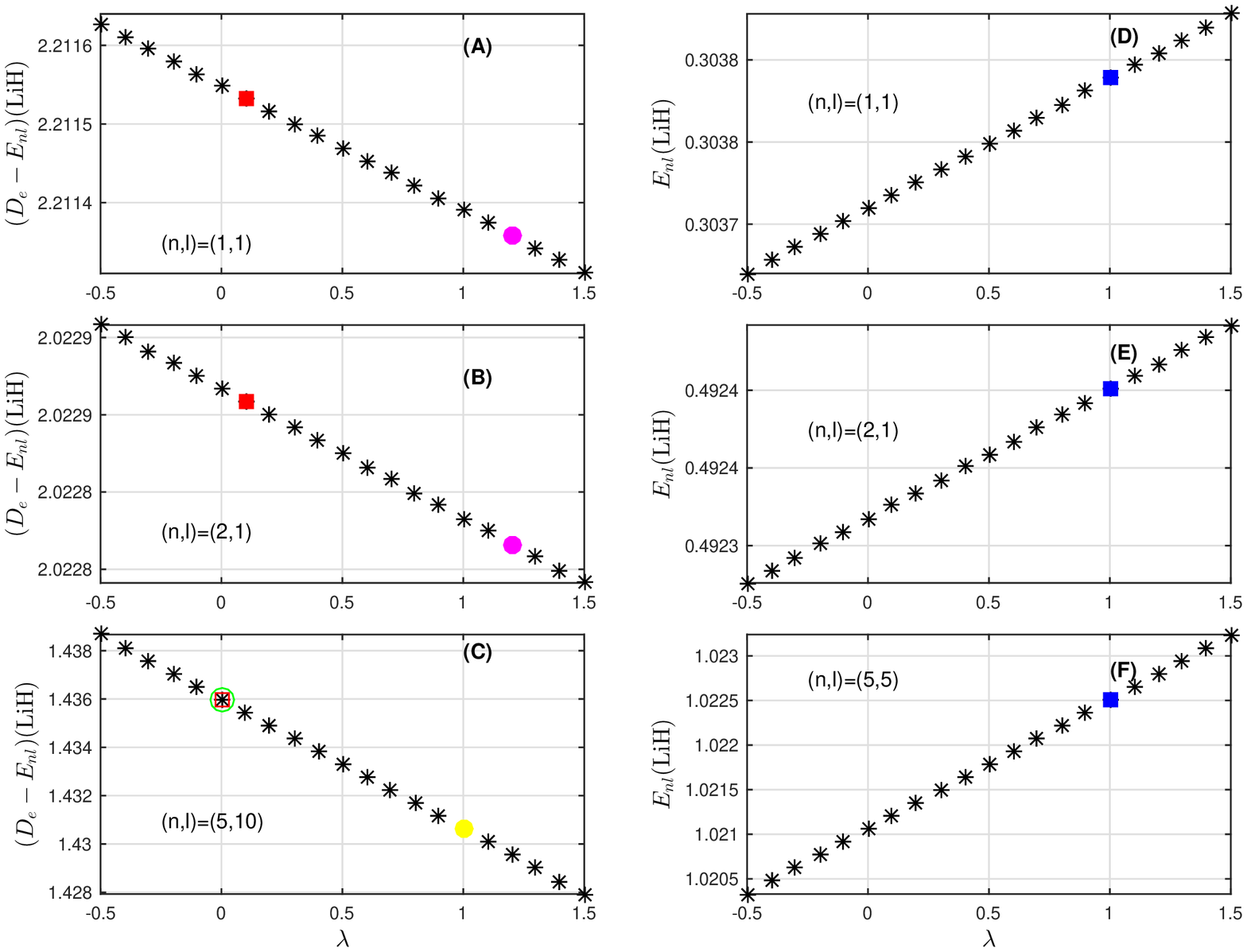}
	\caption{\label{Fig.1energy} Plot of $(D_e-E_{nl})$ (left) and $E_{nl}$ (right) of LiH molecule in eV, against $\lambda$. The state indices are given inside the panels. The colored points show literature energies from various references as: red \cite{roy2014b}, magenta \cite{oluwadare2018}, green \cite{boukabcha2018}, yellow \cite{oyewumi2013a} and blue \cite{oyewumi2013}.}
\end{figure}
\section{Thermodynamical properties in DF potential}
Now we move on to the derivation of thermodynamic quantities using the approximate solution of Sec.~2. The starting point for such calculations 
is the vibrational partition function. This can be constructed from a direct summation over all possible vibrational energy levels available to the 
system, namely, 
\beq\label{partition}
Q(\b)=\sum\limits_{n=0}^{N_{max}} e^{-\b E_{nl}}
\eeq 
where $\b=\f{1}{kT}$, $k$ denotes the Boltzmann constant and $T$ is the temperature. For a finite summation with an upper bound $N$, the Poisson 
summation formula can be written as \cite{strekalov2007},
\beq\label{sum}
\sum\limits_{n=0}^Nf(n)=\f{1}{2}\left[f(0)-f(N+1)\right]+\sum\limits_{m=-\infty}^{\infty}\ds\int_0^{N+1}f(x)e^{-i2\pi mx}dx.
\eeq 
Under the lowest-order approximation $(m=0)$, the above summation formula leads to \cite{strekalov2007},
\beq\label{sum.m0}
\sum\limits_{n=0}^Nf(n)=\f{1}{2}\left[f(0)-f(N+1)\right]+\ds\int_0^{N+1}f(x)dx.
\eeq 
With the help of above relation, one can rewrite Eq.~(\ref{partition}) in following form (`cl' stands for classical),
\beq
\ba{lr}
Z=Q_{cl}(\b)
&=e^{-\b\,d_4}\left\{\f{1}{2}e^{\f{\a^2\hbar^2\b}{2\mu}\left(\f{d_3^2}{L^2}+\f{L^2}{4}\right)}-\f{1}{2}e^{\f{\a^2\hbar^2\b}{2\mu}\left(\f{d_3^2}{(N_{\max}+L)^2}+\f{(N_{\max}+L)^2}{4}\right)}+J^{(0)}_0(\b)\right\} 
\ea 
\eeq 
where
\beq
\ba{ll}
J^{(j)}_{m}(\b)&=\ds\int_{L}^{N_{max}+L}\left(\f{d_3^2}{x^2}+\f{x^2}{4}\right)^j\exp\left[\f{\a^2\hbar^2\b}{2\mu}\left(\f{d_3^2}{x^2}+\f{x^2}{4}\right)-i2\pi mx\right]\,dx\\
N_{max}&=\left[\sqrt{2\,|d_3|}-L\right]+1.
\ea
\eeq 
$N_{max}$ is a (+)ve integer, $\left[x\right]$ denotes the integral part of $x$, and $\left[x\right]\le x$. The energy is maximum at level $n=N_{max}$. The total partition function is a sum of classical $(m = 0)$ and quantum corrected ($Q_{QC}$) contribution corresponding 
to $(m\ne 0)$, i.e., 
\beq
Q_{T}(\b)=Z(\b)+Q_{QC}(\b)
\eeq
where
\beq
Q_{QC}(\b)=\ds \exp\left[-\b\,d_4\right]\sum\limits_{\substack{m=-N\\m\ne 0}}^{N}\,\exp\left[i2\pi m L\right]J^{(0)}_{m}(\b)
\eeq
Now one can proceed for evaluation of quantities like internal energy $(U_T(\b,N), U_{cl}(\b))$, specific heat $(C_T(\b,N), C_{cl}(\b))$, 
free energy $(F_T(\b,N), F_{cl}(\b))$ and entropy $(S_T(\b,N), S_{cl}(\b))$ by using the usual expressions. The total partition function and 
corresponding thermal properties can be conveniently expressed in following simplified manner, 
\beq
\ba{lll}\label{UCFS.T}
Q_T(\b,N)&=H_0(\b,N),              &        Z=Q_{cl}=H_0(\b,0)  \\
U_T(\b,N)&=-\ds\f{1}{Q_T}\f{\partial Q_T}{\partial\b}=d_4-\f{H_1}{Q_T},  &     U_{cl}(\b)=U_T(\b,0)          \\
C_T(\b,N)&=-k_{\b}\b^2\f{\partial U}{\partial\b}
=\ds k_{\b}\b^2\left(\f{H_2}{Q_T}-\f{H_1^2}{Q_T^2}\right), &       C_{cl}(\b)=C_T(\b,0)      \\
F_T(\b,N)&=-k_{\b}T\ln Q_T=-\f{1}{\b}\ln Q_T,  &   F_{cl}(\b)=F_T(\b,0)                  \\
S_T(\b,N)&=k_{\b}\ln Q_T+k_{\b}T\f{\partial \ln Q_T}{\partial T}=k_{\b}\ln Q_T+k_{\b}T\left(\f{H_1}{Q_T}-d_4\right), &  S_{cl}(\b)=S_T(\b,0)
\ea
\eeq 
where
\beq
\ba{ll}
H_j(\b,N)& =\left(\f{\a^2\hbar^2}{2\mu}\right)^j e^{-\b\,d_4} \left[ \f{1}{2}\left(\f{L^2}{4}+\f{d_3^2}{L^2}\right)^j 
e^{\f{\a^2\hbar^2\b}{2\mu}\left(\f{L^2}{4}+\f{d_3^2}{L^2}\right)} - \f{1}{2}\left(\f{(N_{max}+L)^2}{4}+\f{d_3^2}{(N_{max}+L)^2}\right)^j 
\right. \\
&  e^{\f{\a^2\hbar^2\b}{2\mu}\left(\f{(N_{max}+L)^2}{4}+\f{d_3^2}{(N_{max}+L)^2}\right)} 
\left.  +\sum\limits_{\substack{m=-N}}^{N}\,\exp\left[i2\pi m L\right]J^{(j)}_{m}(\b)\right], \\
\f{\partial H_j(\b,N)}{\partial \b}&=H_{j+1}(\b,N)-d_4\,H_j(\b,N).
\ea
\eeq 
The classical partition function and the corresponding classical thermal quantities are given in the right side. 

\begin{figure} [htp] 
	\centering
	\includegraphics[width=16cm,height=18cm]{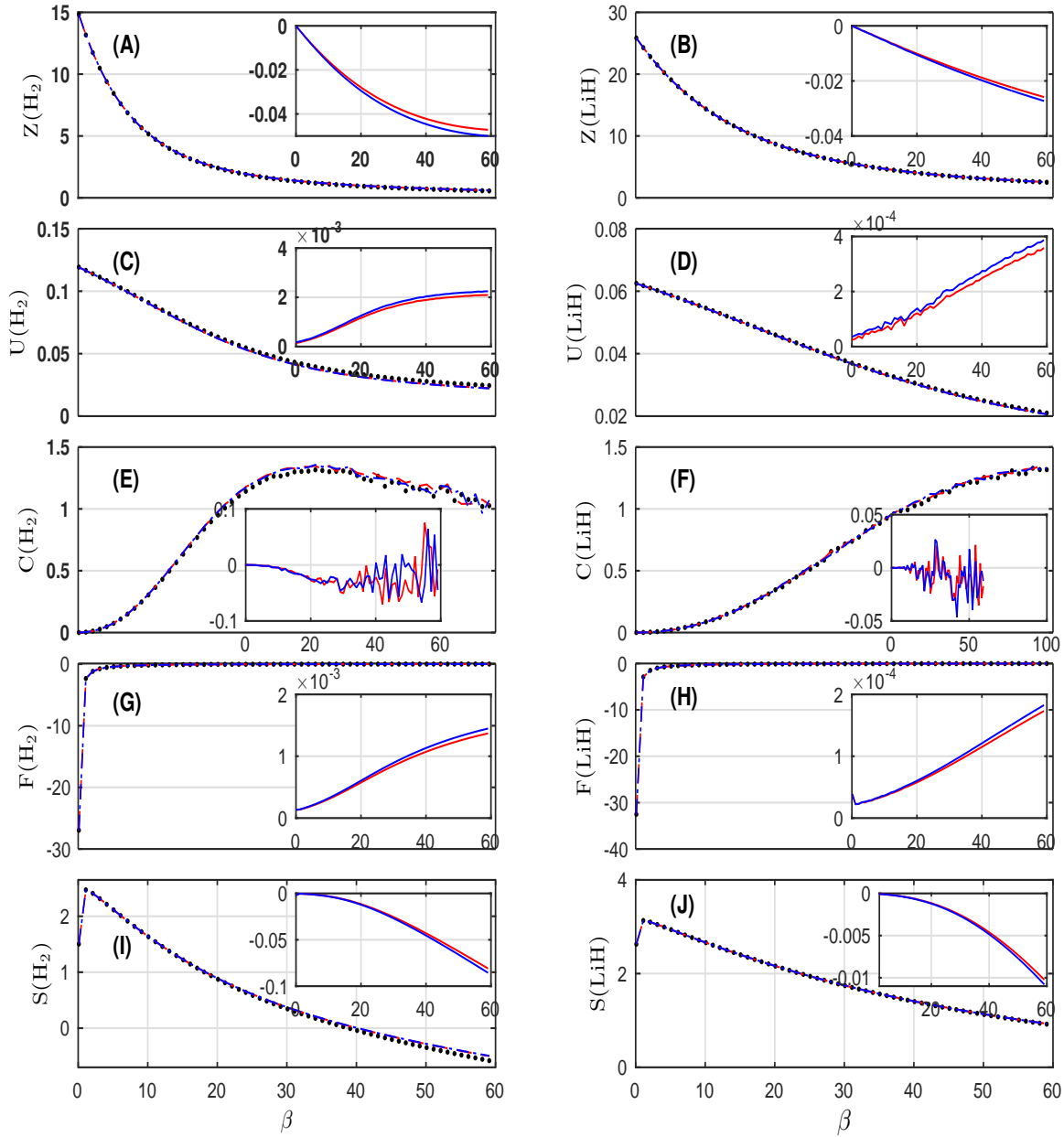}
	\caption{\label{Fig.4thermal} Plot of classical and $N$-corrected thermal properties as functions of $\beta$, for H$_2$ (left) and LiH (right), in atomic unit. The classical quantities are shown in black, while the effect of quantum correction on thermal properties are shown for $N=5$ (red) and $N=10$ (blue). Inset of each figure shows the difference between classical and quantum corrected thermal properties. Insets of (A), (B) represent $(Q_{cl}(\b)-Q_T(\b,5))$ (red) and $(Q_{cl}(\b)-Q_T(\b,10))$ (blue) for H$_2$ and LiH. In a similar manner, other quantities are depicted as follows: (C),(D): internal energies, (E),(F): specific heats, (G),(H): free energies, and (I),(J): entropies.}
\end{figure}

Figure 4 depicts various calculated quantities, such as $Z, U, C, F, S$, for H$_2$ and LiH in left and right panels respectively, with respect to 
$\beta$, in atomic unit. The classical partition function $Q_{cl} (\beta)$ along with its 5th-order quantum correction, $Q_T(\beta, 5)$ and 10th 
order, $Q_T(\beta, 10)$. All these calculations are performed with energies corresponding to $\lambda=-0.25$ in Eq.~(34), while $N_{max}$ for 
a particular molecule is given in Eq.~(45). They all monotonically increase (decrease) with temperature ($\beta$). The differences in $Z$ from 
the classical and quantum-corrected counterparts are shown in the onset panels in red and blue 
colors for 5th and 10th orders respectively. It is seen that, the two differences tend to converge in the low-$\beta$ region and show some 
deviations for large $\beta$. In other words, the $m \neq 0$ terms in Eq.~(42) should be taken in to consideration for vibrational partition 
function in the low temperature region. Next, panels (C), (D) illustrate the changes in vibrational mean energies in H$_2$ and LiH, along with the 
effects of quantum correction inside. In both cases, $U$ fall off with $\beta$ starting from a certain positive value at $\beta \to $0, i.e., 
it grows monotonically with T. There is a slight difference in the nature of plots for two molecules; for LiH it is almost straight, but for 
H$_2$, one finds a minor bent. However, like 
the plots in (A), (B), the differences in this occasion also remain small; and as $\beta$ progresses, the 5th and 10th order corrections 
show some variations (somehow rather more pronounced for LiH). Slight fluctuation is observed in case of LiH. For H$_2$, the curve shows some 
flattish character at low T. The dependence of vibrational specific heat $U$ for the two molecules are now recorded in segments (E) and (F). 
At first it gradually increases with $\beta$ to reach a maximum and thereafter tends to decline. In both molecules, the quantum corrections 
make significant impacts in the behavior of $C$ in low T region, with vigorous fluctuations in the corresponding differences. The vibrational 
free energies $F$ are produced in (G) and (H), which show a sharp rise at low $\beta$ to reach a maximum value, and then approaching a constant 
with further growth in $\beta$. The differences in classical and quantum corrections in this case appear to be rather small, while the 5th and 
10th order results remain in harmony with each other. The last two panels (I) and (J) depict the analogous plots for vibrational entropy. 
After attaining a sharp maximum at lower $\beta$, for both molecules it monotonically decay. Similar plots are given in Fig.~(5) for HCl and 
CO in left and right sides. The vibrational partition functions drop sharply in low $\beta$ region in contrast to the previous figure. The 
differences in $Z$ from classical to $N$th corrected quantity apparently become more significant in CO. The features in $U$ plots remain 
very similar. One notices much stronger deviations from classically and corrected $C$ in case of HCl amongst all the four molecules 
considered in these two figures. Also for CO, the variation of $C$ differs from the rest of molecules. The $F$ plots in these molecules 
resemble those of previous figure. The vibrational entropies very sharply decay at low $\beta$ and thereafter remains practically constant 
throughout the whole range of $\beta$ studied. 

\begin{figure} [htp] 
	\centering
	\includegraphics[width=18cm,height=20cm]{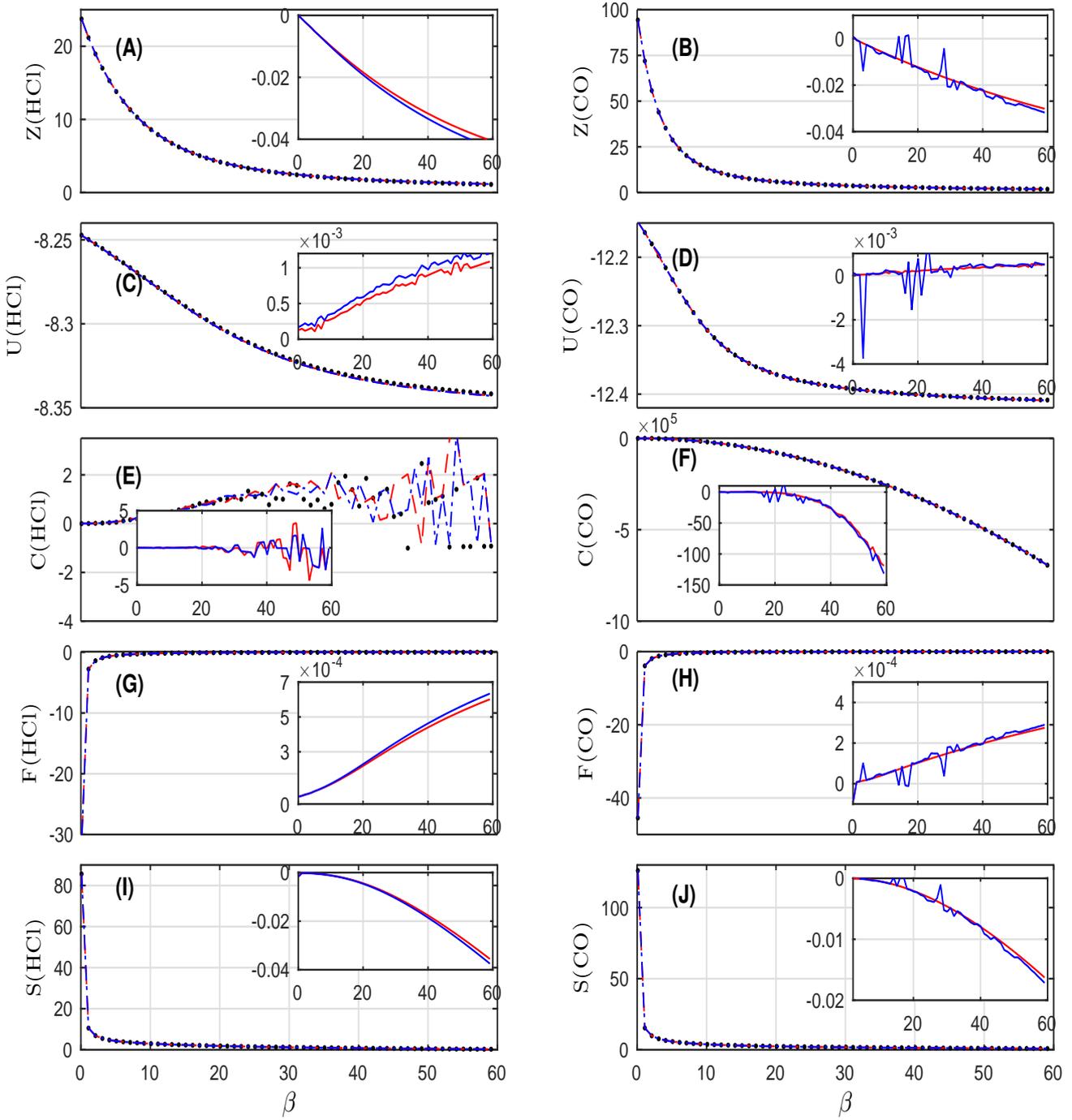}
	\caption{\label{Fig.5thermal} Plots of $Z, U, C, F, S$ as functions of $\beta$, for HCl (left) and CO (right), in atomic unit, as in Fig.~(4).}
\end{figure}

Finally, to estimate the role of parameter $\lambda$ on the thermodynamical quantities, Fig.~(6) portrays the relevant differences in $Z, 
U, C, F, S$ for H$_2$ and LiH molecules in left and right panels respectively. In this case we employ classical partition functions for 
all the calculations. In each case the deviations of corresponding quantities are calculated with respect to $\lambda=0$. Five $\lambda$ are
chosen for this, \emph{viz.}, $-$0.25, 0.25, 0.75, 1 and 1.25. It is observed that, the differences are in general, not very significant. 
One finds the impact to be small through out the whole range of $\beta$ undertaken here. 

\begin{figure} [htp] 
	\centering
	\includegraphics[width=18cm,height=16cm]{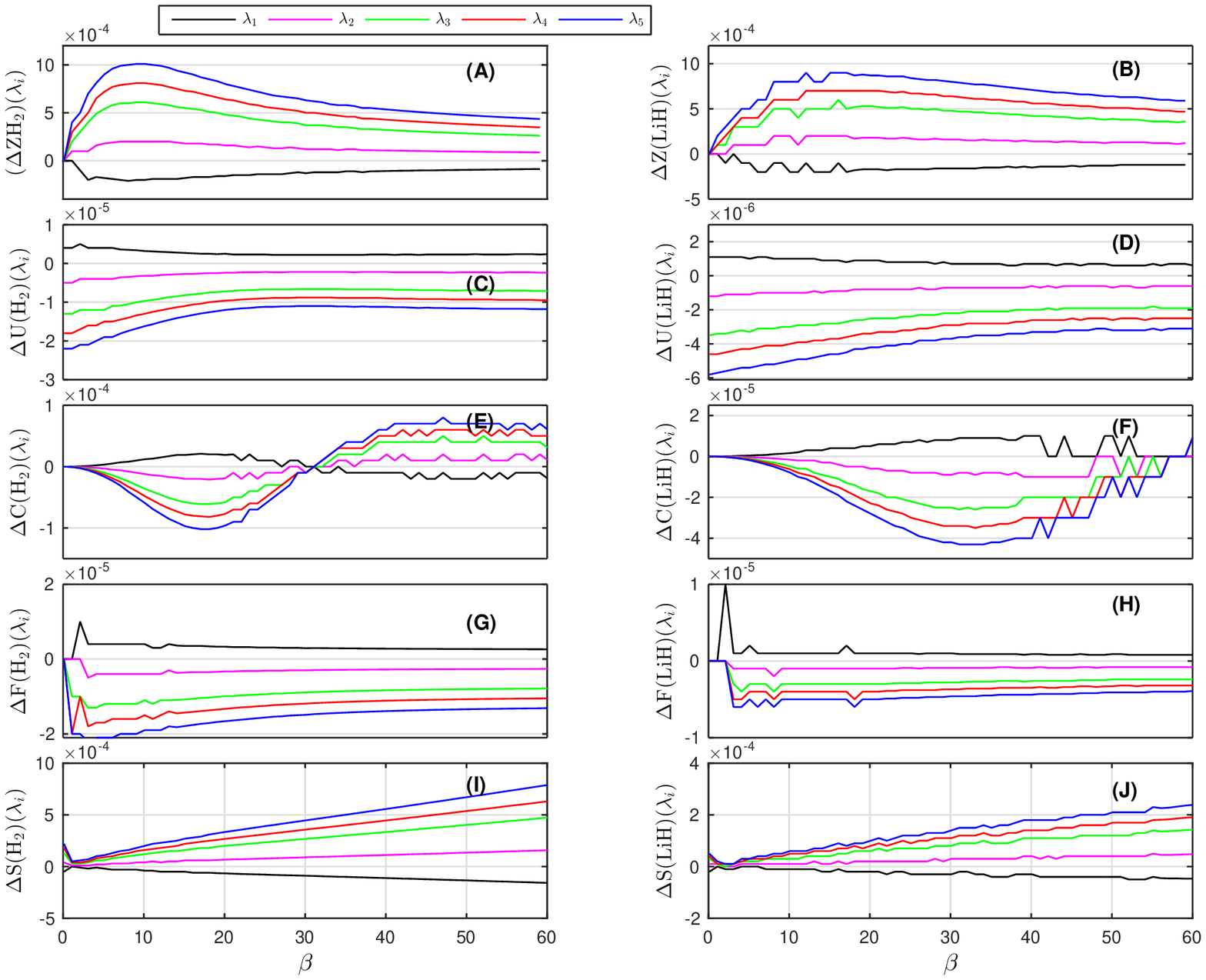}
	\caption{\label{Fig.3Effect.thermal} The effect of $\lam$ on thermal properties are plotted for H$_2$ (left) and LiH (right) in atomic unit. Following differences are given: classical partition functions in (A)-(B), internal energies in (C)-(D), specific heats in (E)-(F), free energies in (G)-(H), and entropies in (I)-(J). They are defined as: $\Delta Z(H_2)(\lam_i)=\left[Z(\b)\right]_{\lam=0}-\left[Z(\b)\right]_{\lam=\lam_i}$, where $\lam_1, \lam_2, \lam_3, \lam_4, \lam_5$ correspond to $-$0.25, 0.25, 0.75, 1 and 1.25. Other quantities follow accordingly.}
\end{figure}

\section{Conclusion}
A simple useful approximation for the centrifugal term is proposed for solution of Schr\"odinger equation in certain potentials through an 
amalgamation of $r \to0$ and {\bf $r \to r_e$ limits}. This is applied to derive expressions for eigenvalues and eigenfunctions in Deng-Fan molecular 
potential. The method offers accurate results ro-vibrational energies for $l =0$ as well as $l \neq 0$ states. This is tested for four molecules 
(H$_2$, LiH, HCl, CO) through a comparison with other existing results in the literature. The effect of $\lambda$ on energies is analyzed. These 
are now utilized to calculate the vibrational partition function and thermodynamic properties such as internal energy, specific heat, free energy 
and entropy in two representative diatomic molecules (H$_2$, LiH). The impact of quantum correction as well as $\lambda$ on these properties is 
discussed by considering the quantum correction up to 10th order, which suffices our purpose. The method can be easily extended to other 
potentials (both central and non-central) including molecular potentials, such as Manning-Rosen, P\"oschl-Teller, H\"ulthen, etc. It would be 
worthwhile to verify its applicability and performance in these cases, some of which may be taken up in future.  

\section {Acknowledgement}
AKR gratefully acknowledges financial support from MATRICS, DST-SERB, New Delhi (sanction order: MTR/2019/000012).   \\ \\


\end{document}